\documentclass[twocolumn,showpacs,preprintnumbers,amsmath,amssymb]{revtex4}

\usepackage{graphicx}
\usepackage{dcolumn}
\usepackage{bm}

\begin{document}


\title{Properties of a fractional derivative Schr\"odinger type wave equation and a new interpretation of
                                       the charmonium spectrum}

\author{Richard Herrmann}
\affiliation{%
GigaHedron, Farnweg 71, D-63225 Langen, Germany\\
}%
\date{\today}
\begin{abstract}
Based on the Caputo fractional derivative the classical, non relativistic  Hamiltonian is quantized leading to 
a fractional  Schr\"odinger type wave equation. The free particle solutions are localized  in space.
Solutions for the infinite potential well and the radial symmetric ground state solution are 
presented. It is shown, that the behaviour of these functions may be reproduced by an ordinary
Schr\"odinger equation with an additional potential, which is of the form $V \sim |x|$ for $\alpha<1$, 
corresponding to the confinement potential, which is introduced phenomenologically to the standard models
for a non relativistic interpretation of quarkonium-spectra. The ordinary Schr\"odinger equation is triple
factorized and yields a fractional wave equation with internal $SU(3)$ symmetry. The twofold iterated version
of this wave equation shows a direct analogy to the derived fractional Schr\"odinger equation.
The angular momentum eigenvalues are calculated algebraically. 
The resulting mass formula is applied to the charmonium spectrum and reproduces the experimental masses
with an accuracy better than $0.1 \% $. Extending the standard charmonium spectrum, three additional particles
are predicted and associated with $\Sigma_c^0(2455)$ and $Y(4260)$ observed recently and one $X(4965)$, not yet 
observed.
The root mean square radius for $\Sigma_c^0(2455)$  is calculated to be $<\!r\!> \approx 0.3 [\text{fm}]$. 
The derived results indicate, that a fractional wave equation may be an appropriate tool for a description
of quark-like particles.
\end{abstract}
\maketitle
\section{Introduction}
Since Newton{\cite{newton}}
 and Leibniz{\cite{leibniz}}
 introduced the concept of infinitesimal calculus,
differentiating a function $f(x^1,...,x^n)$ with respect to the variable $x^i$ is a standard 
technique applied in all branches of physics. The derivative operator $\partial_i$,
\begin{equation} 
\partial_i = \frac{\partial}{\partial x^i}
\end{equation} 
transforms like a vector, its contraction yields the Laplace-operator, using Einstein's sum convention
\begin{equation} 
\Delta = \partial^i \partial_i
\end{equation} 
which is a second order derivative operator,   the essential contribution to establish a wave equation, 
which is the starting point to describe several kinds of wave phenomena.

Until now, in high energy physics the derivative operator has only been used in integer steps. We want to extend the idea
of differentation to arbitrary, not necessarily integer steps.
A natural generalization is to search for an operator $D_i$ by setting
\begin{equation} 
D_i^m = \partial_i^n
\end{equation} 
where $m,n$ are integers. Formally, this is solved by extracting the $m$-th root
\begin{equation} 
D_i = \partial_i^{n/m} \qquad\qquad\qquad\qquad m,n \in \mathbb{N}
\end{equation} 
or, even more general, we will introduce a fractional derivative operator by
\begin{equation}
\label{d} 
D_i = \partial_i^\alpha      \qquad\qquad\qquad\qquad\quad \alpha \in \mathbb{R}_+
\end{equation} 
with the fractional derivative coefficient $\alpha$ being a positive, real number.

The concept of fractional calculus has stimulated mathematicians since the days 
of Leibniz{\cite{f1}}-{\cite{riemann}}.
In physics, early attempts in the 
field of applications was studies on non-local dynamics, e.g. anomalous 
diffusion or fractional Brownian motion {\cite{f3}},{\cite{f4}}.

During the last decade, remarkable progress has been made in the theory of fractional 
wave equations{\cite{raspini}}-{\cite{laskin}}.
Raspini{\cite{raspini}},{\cite{raspini1}} has derived a fractional ($\alpha=2/3$) Dirac equation. Baleanu{\cite{baleanu}} has studied the
Euler-Lagrange equations for classical fields and gave the explicit form of a fractional Klein-Gordon-equation
and a fractional Dirac-equation, conformal with Raspini's.

Both studies were based on the use of the Riemann-Liouville (RL) fractional derivative, which 
is used by many authors working on the field of fractional derivatives.

For practical purposes, the main deficiency of the RL fractional derivative is the fact, that the derivative of 
a constant function does not vanish.
Maybe this is one reason for the fact, that until now, there exists not a single application in the area of
high energy physics.

Laskin{\cite{laskin}} has proposed a hermitean fractional Schr\"odinger equation, based on Feynman's path integral approach.
His applications are based on the semi classical Bohr-Sommerfeld quantization condition only.

We will use a different approach. We will apply the concept of fractional derivative to derive a fractional
Schr\"odinger type wave equation by a quantization of the classical non relativistic  Hamiltonian.
We will collect arguments and results which indicate, that this equation is an alternative
 tool for a appropriate description of the charmonium spectrum, which is normally described
by a phenomenological potential.

In the following sections, we will explicitely derive exact solutions for the free particle and
for particles in an infinite well potential. We will prove, that these solutions show a behaviour,
which may be reproduced by an ordinary Schr\"odinger equation with an additional linear potential term
for $\alpha<1$, indicating that a fractional wave equation and the confinement problem are strongly related.

We will derive a fractional multi-component wave equation via threefold factorization of the ordinary
non relativistic Sch\"odinger equation, which contains an internal $SU(3)$ symmetry.

We will then study an analytical mass formula in terms of angular momentum multiplets which will 
reproduce the experimental masses of the charmonium spectrum within an error of better than $0.1\%$.

We will
extend the standard charmonium spectrum and predict new, additional  particles.
 
Finally, we will give a reasonable estimate for the root mean square radius of $\Sigma_c^0(2455)$.

\section{Fractional derivative}

Let $q = [\alpha]$ be  the integer part of $\alpha$ and $f$ a function of n variables $x^i,\, i=1,...,n$ with $x^i>0$. 
To derive  a specific representation of the fractional derivative operator $D_i$, defined by (\ref{d}) we start with the 
Cauchy integral $^+I_i^\alpha      $ extended to fractional order
\begin{eqnarray} 
^+I_i^\alpha(x^i)f(x^1,...,x^i,...,x^n) &=& \qquad    \\ \nonumber
 \frac{1}{\Gamma(\alpha)} 
     \int_0^{x^i}  du(x^i-u)^{\alpha-1}  f(x^1,...,u,&&...,x^n)
\end{eqnarray} 
a formal split of the partial differential operator into a fractional integral and an integer differential part
\begin{eqnarray} 
\partial_i^\alpha &=& \partial_i^{\alpha -(q+1)} \partial_i^{q+1} \\ \nonumber
         &=& I_i^{(q+1)-\alpha } \partial_i^{q+1} 
\end{eqnarray} 
leads to the definition of the Caputo{\cite{caputo}} fractional differential operator $^{+}_{c}D$, which is the form we will use.
\begin{eqnarray} 
\label{caputo}
^+_cD_i(x^i)f(x^1,...,x^i,...,x^n) = \qquad\qquad&& \\ \nonumber
\frac{1}{\Gamma(q + 1 -\alpha)} 
     \int_0^{x^i}  \frac{du}{(x^i-u)^{\alpha-q}} \frac{\partial^{q+1}}{\partial u^{q+1}}  f(x^1,...,u,& &   ...,x^n)
\end{eqnarray} 
For a constant function this fractional derivative  vanishes:   
\begin{equation} 
\label{const}
^{+}_{c}D(x)\, \text{const} =0
\end{equation} 
For a function of the type 
\begin{equation} 
f(x) = x^{n \alpha}   \qquad\qquad\qquad\qquad\quad  n \in \mathbb{N}
\end{equation} 
the fractional derivative is:
\begin{equation}
^{+}_{c}D(x)\,  x^{n \alpha} = \frac{\Gamma(1+n \alpha)}{\Gamma(1 + (n-1)\alpha)}x^{(n-1) \alpha} 
\end{equation}
For $ k>0,x>0 $ we are then able to define Caputo-Taylor series of the form
\begin{equation}
f(k x) = \sum_{n=0}^\infty a_n (k x)^{n \alpha} 
\end{equation}
The corresponding fractional  derivatives are given by:
\begin{equation}
^{+}_{c}D f(k x) = k^\alpha  \sum_{n=0}^\infty a_{n+1} \frac{\Gamma(1+(n+1) \alpha)}{\Gamma(1+n \alpha)} (k x)^{n \alpha} 
\end{equation}
Since we intend to use $x$ and the fractional derivative $^{+}_{c}D(x)$  on $\mathbb{R}$, 
the next step is an extension of our definition of 
$x$ and  $^{+}_{c}D(x)$ to negative reals $\mathbb{R}_-$. 

We propose the 
following mappings for $x \rightarrow \bar{\chi}(x)$ and 
$^{+}_{c}D(x) \rightarrow  \bar{D}(x) $:
\begin{eqnarray}
\label{defs} 
\bar{\chi}(x) &=& \text{sign}(x) \,  |x|^\alpha \\
\label{defs2}
\bar{D}(x) &=& \text{sign}(x) \, ^{+}_{c}D(|x|)
\end{eqnarray} 
Besides a unique mapping from $\mathbb{R}_+$ to $\mathbb{R}_+$ and $\mathbb{R}_-$ to $\mathbb{R}_-$ the behaviour under parity transformations 
$\Pi$ is well defined:
\begin{eqnarray}
\Pi \bar{\chi}(x) &=& -\bar{\chi} (x) \\
\Pi \bar{D}(x)  &=& -\bar{D}(x) 
\end{eqnarray}
With these definitions we are able to define series $f$ on $\mathbb{R}$
\begin{equation}
f(\bar{\chi}(k x)) = \sum_{n=0}^\infty a_n \bar{\chi}^n (k x) 
\end{equation}
with a well defined derivative $\bar{D}$
\begin{equation}
\bar{D} f(\bar{\chi}(k x)) = \text{sign}(k) |k|^\alpha \sum_{n=0}^\infty a_{n+1} \frac{\Gamma(1+(n+1) \alpha)}{\Gamma(1+n \alpha)} \bar{\chi}^n (k x) 
\end{equation}

To construct a Hilbert space on functions $f(\bar{\chi}(k x))$ we first define the
integral operator
\begin{equation}
\int_{-x}^x  du^\alpha = \text{sign}(x) ^+I^\alpha(|x|^\alpha) 
\end{equation}
with the fractional scalar product
\begin{equation}
<f|g> = \int du^\alpha \, f^*(\bar{\chi}(k u)) \, g(\bar{\chi}(k'u))  
\end{equation}
An expectation value $<\!\hat{O}\!>$ of an operator $\hat{O}$ may consequently be defined with  
\begin{equation}
<\!f|\hat{O}|g\!> = \int du^\alpha \, f^*(\bar{\chi}(k u)) \, \hat{O} \, g(\bar{\chi}(k'u))  
\end{equation}
to be
\begin{equation}
\label{expect}
<\!\hat{O}\!> = \frac{<\!f|\hat{O}|g\!>}{<\!f|g\!>}  
\end{equation}
The space coordinates and corresponding derivatives, defined by (\ref{defs}) and (\ref{defs2}), are the basic
input for our derivation of a fractional non relativistic Schr\"odinger type wave equation, the corresponding
angular momentum operators and caculation of expectation values.

\section{\label{sgl}Quantization of the classical Hamiltonian and free particle solutions}
By use of the definitions ({\ref{defs}}),({\ref{defs2}})  for the space coordinate and for the fractional derivative,
we are able to quantize the Hamiltonian of a classical non relativistic 
 particle and solve the corresponding Schr\"odinger equation.

We define the following set of conjugated operators on an euclidean space for $N$ particles in space coordinate
representation:
\begin{eqnarray}
\hat{P}_\mu &=& \{  \hat{P}_0,\hat{P}_i\} = \{ i \hbar \partial_t,-i \left( \frac{\hbar}{m c} \right)^{\alpha} m c  \bar{D}_i  \}\\
 &=&  \{ i \hbar \partial_t,-i \left( \frac{\hbar}{m c} \right)^{\alpha} m c \,\, \text{sign}(x^i) \, ^{+}_{c}D(|x^i|)  \}\\
\hat{X}_\mu &=& \{  \hat{X}_0,\hat{X}_i\} = \{ t, \left( \frac{\hbar}{m c} \right)^{(1-\alpha)} \frac{\bar{\chi}(x_i)} {\Gamma(\alpha+1)} \} \\
 &=& \{ t, \left( \frac{\hbar}{m c} \right)^{(1-\alpha)} \!\!\! \!\!\! \frac{1} {\Gamma(\alpha+1)}\, \text{sign}(x_i)|x_i|^\alpha \} \\
& & \qquad\qquad\qquad\qquad\qquad i = 1,..., 3 N \nonumber
\end{eqnarray}
These operators  satisfy the following commutator relations on a function set $\{  x^{n \alpha}  \}$: 
\begin{eqnarray}
\left[ \hat{X}_i, \hat{X}_j \right] &=& 0 \\ 
\left[ \hat{P}_i \, , \hat{P}_j \, \right] &=& 0 \\ 
\label{comm}
\left[ \hat{X}_i, \hat{P}_j \right] &=& -i \hbar \delta_{ij} \frac{1}{\Gamma(1 + \alpha)} \times \\ \nonumber
& & \left( 
 \frac{\Gamma(1 + n \alpha)}{\Gamma(1 + (n-1) \alpha)} -  \frac{\Gamma(1 + (n+1) \alpha))}{\Gamma(1 + n \alpha)}\right) \\ 
&=& -i \hbar \delta_{ij} \, c(n,\alpha)
\end{eqnarray}
With these operators, the classical, non relativistic Hamilton function $H_c$, which depends on
the classical momenta and coordinates $\{p_i,x^i\}$
\begin{equation}
H_c = \sum_{i=1}^{3N}\frac{p^2_i}{2m} +V(x^1,...,x^i,...,x^{3N}) 
\end{equation}
is quantized. This yields the Hamiltonian $H^\alpha$ 
\begin{equation}
H^\alpha = -\frac{1}{2} m c^2 \left( \frac{\hbar}{m c} \right)^{2 \alpha} \bar{D}^i \bar{D}_i +V(\hat{X}^1,...,\hat{X}^i,...,\hat{X}^{3N})
\end{equation}
Thus, a time dependent Schr\"odinger type equation for fractional derivative 
operators results
\begin{eqnarray}
\label{schroedinger}
H^\alpha \Psi &=& (-\frac{1}{2} m c^2 \left( \frac{\hbar}{m c} \right)^{2 \alpha} \bar{D}^i \bar{D}_i\\ 
&+&V(\hat{X}^1,...,\hat{X}^i,...,\hat{X}^{3N})) \Psi = i \hbar \partial_t \Psi \nonumber
\end{eqnarray}
For $\alpha=1$ this reduces to the classical Schr\"odinger equation.

\subsection{Properties of the momentum operator}

We extend the standard series expansion of the exponential function to the fractional case
\begin{eqnarray}
\exp(\alpha,\bar{\chi}(k x)) &=& \sum_{n=0}^\infty \frac{\text{sign}^n (k x) |k x|^{\alpha n}}{\Gamma(1+\alpha n)}  \\ \nonumber
  = \sum_{n=0}^\infty \frac{|k x|^{2 \alpha n}}{\Gamma(1+2 \alpha n)} &+&  \text{sign}(k x) \sum_{n=0}^\infty \frac{ |k x|^{(2 n + 1)\alpha }}{\Gamma(1+(2 n + 1)\alpha)} \\
 &=& \mathfrak{E}_\alpha( \bar{\chi}(k x))
\end{eqnarray}
where $\mathfrak{E_\alpha}$ is the Mittag-Leffler function{\cite{ml}}.
The functions $\psi = \exp(\alpha,-i \bar{\chi}(k x))$ are eigenfunctions of the 
momentum operator with the real eigenvalues
\begin{equation}
\hat{P} \psi = 
\left( \frac{\hbar}{m c} \right)^{\alpha} m c \, \text{sign}(k) \, |k|^\alpha \,  \psi
\end{equation}

The Leibniz product rule, which plays an important role for the standard derivative, 
 is not valid any more for the fractional 
derivative. Instead, with an arbitrary additional function $R$, we can write:
\begin{equation}
\bar{D} (f g) = (\bar{D} f) g + f (\bar{D} g) + R 
\end{equation}
For the momentum operator it follows
\begin{eqnarray}
\int du^\alpha \, f^* (\hat{P}g)  &=& f g 
- \int du^\alpha \, (\hat{P}f^*) g 
- \int du^\alpha \, R  \nonumber \\ 
  &=& + \int du^\alpha \, (\hat{P}f)^* g 
      - \int du^\alpha \, R 
\end{eqnarray}
Consequently, neither $\hat{P}$ nor the fractional  Schr\"odinger operator with  $H^\alpha$ in 
({\ref{schroedinger}}) are hermitean operators.

A direct consequence is the non orthogonality 
of the calculated eigenfunctions. 

In general, hermitean operators are preferred, since 
their eigenvalues and expectation values are always real.  Nevertheless, eigenvalues for momentum, energy and angular momentum as well
as expectation values derived with the proposed  fractional Schr\"odinger equation
({\ref{schroedinger}}) turn out to be real, as will be demonstrated in the following sections. 

Anyhow, we doubt, that a fractional operator should always be hermitean. A typical example was the expectation
value of the root mean square radius of a free quark. Indeed, any real value would be a contradiction to the 
experiment.

\subsection{Free particle solutions}
\begin{figure}
\begin{center}
\includegraphics[width=75mm,height=103mm]{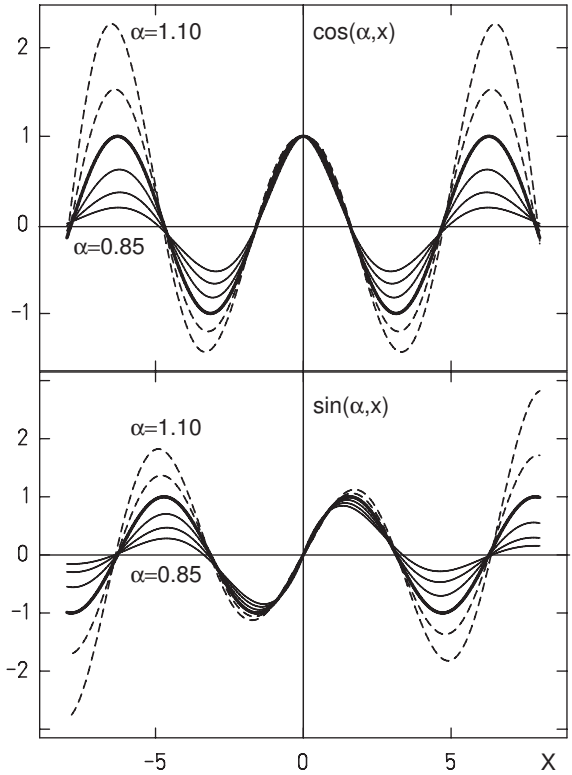}\\
\caption{\label{sincos}Free particle solutions for the fractional derivative operator
Schr\"odinger type equation, for different values of $\alpha$. 
Upper part of figure are $\cos(\alpha,x)$, lower part are $\sin(\alpha,x)$ 
with $\alpha=0.85, 0.90, 0.95$ (solid thin line),  .
$\alpha=1$ (solid thick line) and $\alpha=1.05, 1.10$ (dashed thin lines) each. 
For $\alpha=1$ solutions
reduce to the standard $\cos(x)$ and $\sin(x)$ functions. For $\alpha<1$ these functions are
increasingly located at $x=0$. For $\alpha>1$ the amplitudes increase with 
increasing $x$.} 
\end{center}
\end{figure}
We will now present the free particle solutions for the fractional Schr\"odinger type  equation ({\ref{schroedinger}}). 
We can do this, since 
for $V(\hat{X}^i) = 0$ the commutator $[\hat{P}_\mu,H^\alpha]$ 
vanishes and consequently, energy and momentum are conserved.
\begin{figure}
\begin{center}
\includegraphics[width=76mm,height=103mm]{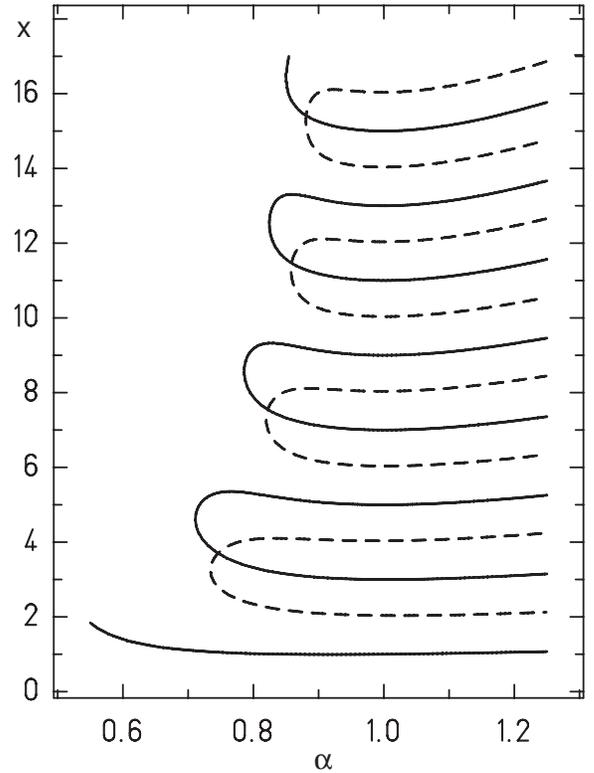}\\
\caption{\label{zero}Zeroes of eigenfunctions $\cos(\alpha,\frac{\pi}{2}x )$ (solid lines) 
and $\sin(\alpha,\frac{\pi}{2}x)$ (dashed lines)
of the free particle eigenfunctions. For $\alpha \leq 1/2$ there are no zeroes. For $\alpha = 1$ the
zeroes are given by $k_n = n \in \mathbb{N}$. For $\alpha \geq 1$ an infinite number of zeroes exists.
} 
\end{center}
\end{figure}

Let us first consider the one dimensional case.  
We extend the definition for the standard series expansion for the sine and cosine function
\begin{eqnarray}
\label{defcossin}
\cos(\alpha,x) &=& \sum_{n=0}^\infty  \frac{(-1)^n \, |x|^{2 n \alpha}}{\Gamma(2n\alpha+1)} \\   
               &=& \mathfrak{E}_{2 \alpha}( -\bar{\chi}^{2} (k x)) \\
\sin(\alpha,x) &=& \text{sign}(x) \sum_{n=0}^\infty  \frac{(-1)^n \,|x|^{(2 n +1)\alpha }}{\Gamma((2n+1)\alpha+1)}  \\
               &=& \bar{\chi}(k x)\, \mathfrak{E}_{2 \alpha, 1+\alpha}( -\bar{\chi}^{2} (k x))
\end{eqnarray}
where $\mathfrak{E}_\alpha$ and $\mathfrak{E}_{\alpha,\beta}$ are Mittag-Leffler{\cite{ml}} 
and generalized Mittag-Leffler functions{\cite{wi}}. With these definitions, the following relations are valid:
\begin{eqnarray}
\bar{D} \sin(\alpha,kx) &= \text{sign}(k)|k|^\alpha \cos(\alpha,kx)  \\
\bar{D} \cos(\alpha,kx) &= -\text{sign}(k)|k|^\alpha \sin(\alpha,kx) 
\end{eqnarray}

It follows from these relations, that the above functions (\ref{defcossin}) are the eigenfunctions of the free
Schr\"odinger type equation ({\ref{schroedinger}}) in one dimension.
In the stationary case we get the energy relation
\begin{equation}
E = \frac{1}{2} m c^2 \left( \frac{\hbar |k|}{m c} \right)^{2 \alpha} 
\end{equation}
This result may easily be extended to the n-dimensional case.

In figure {\ref{sincos}} the functions $\sin(\alpha,x)$ and $\cos(\alpha,x)$ are plotted for different values
of $\alpha$. While for $\alpha=1$, these functions reduce to the known $\cos(x)$ and $\sin(x)$,
which are spread over the whole x-region,  
for $\alpha<1$ these functions become more and more located at $x=0$ and oscillations are damped, 
a behaviour, which we
know e.g. from the Airy-functions. For $\alpha>1$ the functions amplitude increases for increasing $x$.
For $\alpha < 1$ these functions are normalizable on $\mathbb{R}$: There exists an upper bound $M$ with
\begin{eqnarray}
\int_{-\infty}^\infty dx^\alpha \cos(\alpha,k x) \cos(\alpha,k'x) &<& M \\
\int_{-\infty}^\infty dx^\alpha \sin(\alpha,k x) \sin(\alpha,k'x) &<& M \\
\int_{-\infty}^\infty dx^\alpha \cos(\alpha,k x) \sin(\alpha,k'x) &=& 0 
\end{eqnarray}
For $\alpha \geq 1$ this integral is not bound any more, instead a box-normalization with a box size 
$L$ much larger than the dimensions of the system considered is proposed.  

\subsection{Particle in an infinite potential well}
\begin{figure}
\begin{center}
\includegraphics[width=79mm,height=208mm]{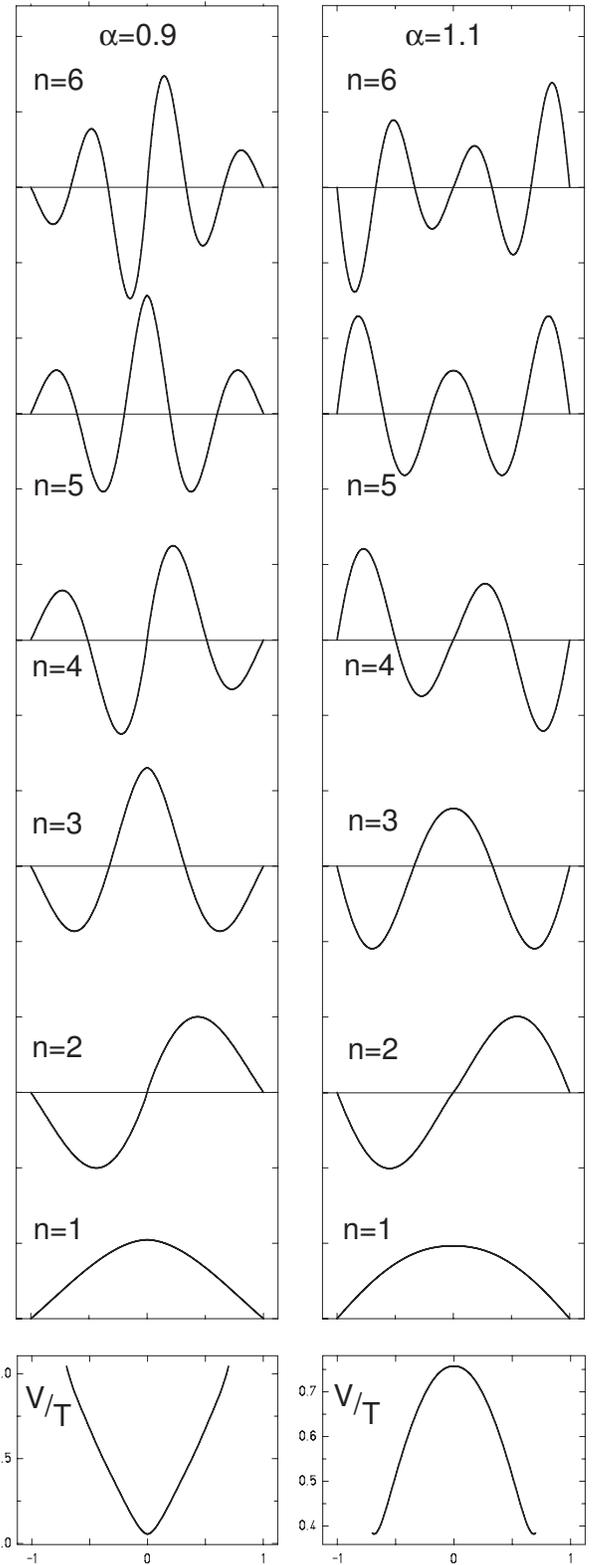}\\
\caption{\label{V} The six lowest eigenfunctions for the one dimensional infinite square well potential, with
$\alpha=0.9$ on the left and $\alpha=1.1$ on the right side. Below the corresponding  potential for
equivalent solutions of the ordinary $(\alpha=1)$ Schr\"odinger equation, according to ({\ref{temperature}}) is plotted.
For $\alpha < 1$ this potential contains a strong linear admixture, for $\alpha>1$ to potential graph behaves like
 $\sim 1-x^2$
} 
\end{center}
\end{figure}
Now we will give the exact eigenfunctions and eigenvalues for a particle confined in an infinite potential well.
We first investigate the one dimensional case. Therefore we define the potential $V(x)$
\begin{equation}
\label{potsquare}
V(x) =
\begin{cases}
0 \qquad \,\,\, |x| \leq a \\
\infty \qquad |x| > a 
\end{cases}
\end{equation}
The corresponding boundary condition for the eigenfunctions $\psi(x)$ is: 
\begin{equation}
\psi(-a) = \psi(a) = 0 
\end{equation}
In figure {\ref{zero}} the zeroes for the free particle solutions $\cos(\alpha,\frac{\pi}{2} x)$ and
 $\sin(\alpha,\frac{\pi}{2} x)$ are plotted. For $\alpha \leq 1/2$ there are no zeroes. For the 
interval $1/2 < \alpha < 1$ exists only a finite set of zeroes. For $\alpha \ge 1 $ an infinite number of zeroes exists.
 
Let $k^0$ be a zero of the free particle solutions. 
The eigenfunctions of the infinite potential well potential are then given by: 
\begin{eqnarray}
\psi_{2 n}^{(+)}(x) &=& \cos(\alpha,\bar{\chi}(k^0_{2 n} \frac{x}{a})) \\
\psi_{2 n+1}^{(-)}(x) &=& \sin(\alpha,\bar{\chi}(k^0_{2 n+1} \frac{x}{a})) 
\end{eqnarray}
where the sign indicates the parity of the states. The normalization condition is
\begin{equation}
\int_{-a}^a dx^\alpha \psi^*_n(x) \psi_n(x)= 1 
\end{equation}
The energy is then given by
\begin{equation}
e_n = \frac{1}{2} \left( \frac{\hbar}{m c}\right)^{2 \alpha} m c^2 |\frac{k^0_n}{a}|^{2 \alpha}
\end{equation}
The extension to the N-dimensional case is then 
\begin{equation}
\Psi_{n_1 n_2 ... n_N} (x^1, x^2, ..., x^N) = \prod_{i=1}^N \psi_{n_i}(x^i)
\end{equation}
and for the energy
\begin{equation}
\label{energy_squarewell}
E_{n_1 n_2 ... n_N} = \frac{1}{2} \left( \frac{\hbar}{m c}\right)^{2 \alpha} m c^2 \sum_{i=1}^N |\frac{k^0_{n_i}}{a_i}|^{2 \alpha}
\end{equation}

\subsection{Radial solutions}

In the case of fractional derivative operators  there exists no general theory of covariant coordinate
transformations until now.

We intend to perform a coordinate transformation from carthesian to 
hyperspherical coordinates in $\mathbb{R}^N$
\begin{equation}
f(x_1,x_2,...,x_N) = f(r, \phi_1, \phi_2,...,\phi_{N-1})
\end{equation}
The invariant line element in the case $\alpha=1$  
\begin{equation}
ds^2 = g_{ij}dx^i dx^j   \quad  i,j= 1,...,N
\end{equation}
for arbitrary fractional derivative coefficient $\alpha$ may be  generalized to 
\begin{equation}
ds^{2\alpha} = g_{ij}^\alpha dx^{i\alpha} dx^{j\alpha}   \quad  i,j= 1,...,N
\end{equation}
Consequently a natural definition of the radial coordinate is given by
\begin{equation}
r^{2\alpha} = \sum_{i=1}^N x_i^{2\alpha}
\end{equation}
We assume the spherical ground state 
to be independent of the angular variables, square integrable 
and of positive parity. Therefore an appropriate ansatz is
\begin{equation}
g(N,\alpha, kr) =  \sum_{n=0}^\infty  (-1)^n  a_n(N,\alpha) (|k|r)^{2\alpha n} 
\end{equation}
or in carthesian coordinates
\begin{equation}
\label{radx}
g(N,\alpha, k x_1,...,k x_N) =  \sum_{n=0}^\infty  (-1)^n  a_n(N,\alpha) (\sum_{i=1}^N |kx|^{2\alpha})^{n} 
\end{equation}
where the coefficients $a_n$ depend on the explicit form of the potential. 

For a free particle, a solution on $\mathbb{R}^N$ is given 
with the abbreviation
\begin{equation}
\eta_j = \Gamma(1+2 \alpha j) / \Gamma( 1 + 2 \alpha (j-1)) 
\end{equation}
by the recurrency relation
\begin{eqnarray}
a_0 &=& 1 \nonumber \\
a_j    & =& a_{j-1}/\left( (N-1)\, j \, \eta_1 + \eta_j \right) \qquad j=1,2,... 
\end{eqnarray}

An infinite spherical well is described by the potential 
\begin{equation}
\label{potsphere}
V(r) =
\begin{cases}
0 \qquad \,\,\, r \leq r_0 \\
\infty \qquad r > r_0 
\end{cases}
\end{equation}
The corresponding boundary condition for the ground state wave function $g(N,\alpha,r)$ is: 
\begin{equation}
g(N,\alpha,r_0) = 0 
\end{equation}
Let $k^0_\text{sph}$ be the first zero of the free particle ground state wave function,
the ground state wave function for the spherical infinite well potential is given by
\begin{equation}
g(N,\alpha,k^0_\text{sph} r / r_0)  
\end{equation}
 and the ground state energy  is then given by 
\begin{equation}
\label{e0}
e_0(N,\alpha)  = \frac{1}{2} m c^2  \left( \frac{\hbar}{m c}\frac{k^0_\text{sph}}{r_0}\right)^{2 \alpha} 
\end{equation}

\subsection{Remarks on equivalent solutions for the ordinary Schr\"odinger equation}
We have shown, that the free particle solutions and the solutions for the infinite
potential square well of the fractional Schr\"odinger equation for $\alpha<1$ are localized at the origin
and for $\alpha>1$ are localized at the boundaries of a given region respectively.

We want to deduce a similar behaviour of these functions in terms of the ordinary $(\alpha=1)$
Schr\"odinger equation.
Let us assume, the eigenfunctions of the fractional Schr\"odinger equation may be 
equivalently
interpreted as solutions of the
ordinary $({\alpha=1})$ Schr\"odinger equation with an additional potential $V$.

In order to derive the explicit form of this potential, we use the following  relation
 between
the eigenfunctions $\Psi_n$, temperature $T$ and a given potential $V$, which is derived  
within the framework of thermodynamics and statistical quantum mechanics{\cite{greiner}}:

Let $\Psi_n$ and $E_n$ be the eigenfunctions and energy eigenvalues of the ordinary Schr\"odinger equation
with a given potential $V$:
\begin{equation}
\left( - \frac{\hbar^2}{2m}\Delta + V \right) \Psi_n = E_n \Psi_n
\end{equation}
As long as the temperature $T$ is large compared to the average level spacing, the relation ($N$ is the normalization constant)
\begin{eqnarray}
\label{temperature}
\frac{1}{\sqrt{N}}\exp(-V/T) &=&
 \frac{\displaystyle \sum_{n=0}^\infty \Psi^*_n \Psi_n \exp(-E_n/T)}{\displaystyle \sum_{n=0}^\infty \exp(-E_n/T)} \\
& & \nonumber \\
&& T \gg \Delta E_n = E_{n+1}-E_n \nonumber 
\end{eqnarray}
is valid. 

Since eigenfunctions and eigenvalues for the fractional free particle solutions are known, we 
therefore are able to deduce the explicit form of such a potential.

In figure \ref{V} the graph of $V/T$ is plotted for $\alpha=0.9$ and $\alpha=1.1$ respectively. For $\alpha=0.9$
the potential contains a dominant linear term $V \sim |x|$  , while 
for  $\alpha=1.1$ a behaviour like $V \sim 1-x^2$ may
be deduced. Of course, for $\alpha=1$ a potential $V = \text{const}$ results.

The behaviour of the fractional eigenfunctions may alternatively be interpreted, assuming that $\alpha$ is a measure
of charge for a particle moving in an homogenously charged background. 

For $\alpha > 1$, we could assume a charged
particle, with charge e.g. $\alpha$ moving in a homogeneous background of charged particles with charge $1-\alpha$.
Since for a homogeneously charged sphere and approximately for a charged box too, the potential inside the box is 
$V \sim -Z(1-r^2)$ with $Z=1-\alpha$. This obviously would explain a repulsive force. The energetically
favoured positions for a particle with charge $\alpha$ indeed were the boundaries of the box. 

For $\alpha=1$ the
background was neutral and therefore no additional interaction with a particle was present.

For $\alpha<1$ this simple model could explain an attraction, but not the linearity of the potential.

Therefore we obtain the remarkable result, that in the case $\alpha<1$, the free particle solutions of
the fractional Schr\"odinger equation show a behaviour, which is equivalent to the behaviour of solutions
of the ordinary Schr\"odinger equation with an additional linear potential term. In other words, the solutions 
of a free fractional
wave equation with $\alpha < 1$ automatically show confinement, a property, which was first observed for quarks.

In order to obtain more properties of the fractional derivative operator Schr\"odinger equation, we will now
calculate the eigenvalues of the angular momentum operator.

\section{Classification of angular momentum eigenstates}
We define the generators of infinitesimal rotations in the $i,j$-plane
 ($i,j=1,...,3N$), with $N$ being the
number of particles):
\begin{eqnarray}
L_{ij} &=& \hat{X}_i \hat{P}_j - \hat{X}_j \hat{P}_i \nonumber \\
       &=& -i \hbar \left( \frac{\bar{\chi}(x_i)}{\Gamma(\alpha+1)}\bar{D}_j -\frac{\bar{\chi}(x_j)}{\Gamma(\alpha+1)}\bar{D}_i \right)
\end{eqnarray}
We will derive the angular momentum eigenvalues algebraically. 
Thus it is necessary, to apply the commutator relation $[\hat{X}_i, \hat{P}_j ] $ (see ({\ref{comm}})) repeatedly.
\begin{figure}
\begin{center}
\includegraphics[width=75mm,height=58mm]{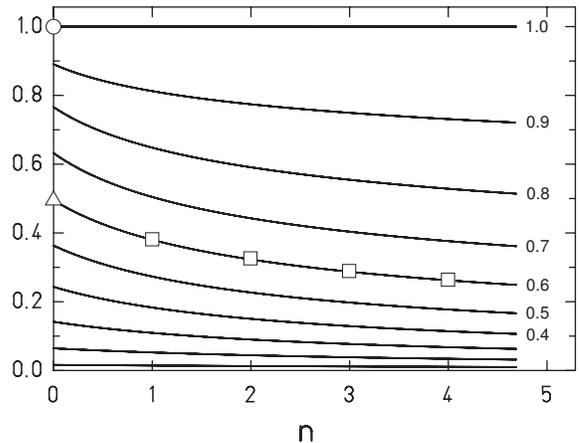}\\
\caption{\label{commutator}Commutator $[\hat{X}_i, \hat{P}_i ] $ from ({\ref{comm}}) in units of $\hbar$ 
on the function set $\{x^{n\alpha}\}$ for
different values of $\alpha$. The dependence on $n$ is a direct consequence of the fact, that the Leibniz 
product rule
is not valid any more for fractional derivatives. For $\alpha=0.6$  
the positions of the successive 
approximations on $c(n,\alpha)$ according to ({\ref{cp1}}),({\ref{cp2}}) and ({\ref{cp3}}) are given as 
circle, triangle and squares.
} 
\end{center}
\end{figure}
In figure {\ref{commutator}}, this commutator $c(n,\alpha)$ is plotted. It shows a smooth
dependence on $n$, which we have to eliminate. 
\begin{table*}
\caption{\label{tabone}Eigenvalues in units of $\hbar$ for the angular momentum states of a single particle.
$n$ is a counter for the eigenvalues of the Euler operator, eigenvalues for $L_z(\alpha)$ and
$J^2_c(\alpha)$ are given for $\alpha=1$, $\alpha=2/3$, $\alpha=0.68$ and $\alpha=0.65$. $J^2_c(\alpha)$ are listed with c  according
to ({\ref{cp1}}),({\ref{cp2}}) and ({\ref{cp3}}). }
\begin{ruledtabular}
\begin{tabular}{@{}*{8}{r}}
                             
$n=L_z(1)$ & $L_z(2/3)$  & $L_z(0.68)$ & $J^2_0(1)$ & $J^2_0(2/3)$ & $J^2_0(0.68)$ & $J^2_1(0.65)$ & $J^2_2(j,0.65)$ \cr
0            &0          & 0            &  0         &  0           &  0            &     0         &   0         \cr
1            &1          & 1            &  2         &  2           &  2            &1.604 767      &   1.478 157\cr
2            &1.460 998  & 1.478 157    &  6         &  3.595 515   &  3.663 108    &3.078 892      &   2.800 590 \cr 
3            &1.860 735  & 1.894 649    & 12         &  5.323 069   &  5.484 346    &4.735 519      &   4.305 776 \cr 
4            &2.222 222  & 2.272 597    & 20         &  7.160 493   &  7.437 298    &6.539 094      &   5.961 779 \cr 
5            &2.556 747  & 2.623 332    & 30         &  9.093 704   &  9.505 205    &8.468 379      &   7.747 796 \cr 
6            &2.870 848  & 2.953 417    & 42         & 11.112 618   & 11.676 094    &10.508 808     &   9.649 033 \cr 
\end{tabular}
\end{ruledtabular}
\end{table*}

Ignoring both, the $n$ and $\alpha$ dependence we set as  a lowest order approximation:
\begin{equation}
\label{cp1}
c_0 = c(0,1)= 1  
\end{equation}
A more precise statement for $c(n,\alpha)$ can be deduced from the following consideration:
Since we will concentrate 
on the lowest enery levels only , the approximation
\begin{equation}
\label{cp2}
c_1(\alpha) = c(0,\alpha) = 1 - \frac{1}{\Gamma(1-\alpha)\Gamma(1+\alpha)} 
\end{equation}
is valid. Of course, this overestimates the commutator for higher values of $n$. Therefore a more 
sophisticated treatment fixes $c(n,\alpha)$ for a given $j$ to be: 
\begin{equation}
\label{cp3}
c_2(j,\alpha) =  
\frac{\Gamma(1 + (j+1)\alpha)}{\Gamma(1 + j\alpha )\Gamma(1 + \alpha)}  -  \frac{\Gamma(1 + j\alpha)}{\Gamma(1 + (j-1)\alpha)\Gamma(1 + \alpha)} 
\end{equation}
We will consider these three cases, which allows an estimate on the validity of the results. 
Therefore, as long as the commutator does not depend on $n$, 
  $[L_{ij},H^\alpha]$ vanishes and angular momentum is conserved. 
Commutator relations for $L_{ij}$ are isomorph to an extended  fractional $SO^\alpha(3N)$ algebra:
\begin{equation}
[ L_{i   j  } ,    L_{m   n    } ] =
 i \hbar \, c(\alpha) \,(
\delta_{i   m   } L_{j   n   } +
\delta_{j   n   } L_{i   m   } -
\delta_{i   n   } L_{j   m   } -
\delta_{j   m   } L_{ i  n   } ) 
\end{equation}
Consequently, we can proceed in a standard way {\cite{tb}}, 
by defining the Casimir operators
\begin{equation}
\Lambda_k^2=\frac{1}{2} \sum_{i,j}^{k} (L_{ij})^2 \qquad, \qquad k=2,...,3N
\end{equation}
which indeed fulfill the relations 
$[\Lambda_{3N}^2, L_{ij} ] = 0$
and successively
$[\Lambda_k^2, \Lambda_{k'}^2 ] = 0$.
Consequently the successive group chain 
\begin{equation}
SO^\alpha(3 N) \supset
SO^\alpha(3 N-1) \supset  \ldots
        \supset
SO^\alpha(3) \supset
SO^\alpha(2)
\end{equation}
is established.
The explicit form of the Casimir operators is given by
\begin{equation}
\Lambda_k^2 = 
+ \hat{X}^i \hat{X}_i \hat{P}^j \hat{P}_j - i \hbar c(\alpha)\,(\delta_i^i-1) \hat{X}^j \hat{P}_j - \hat{X}^i \hat{X}^j \hat{P}_i \hat{P}_j
\end{equation}
We introduce a generalization of the homogeneous Euler operator
for fractional derivative operators 
\begin{equation}
J_e^k(\alpha)     = \bar{\chi}(x^i)/\Gamma(1+\alpha) \, \bar{D}_i
\end{equation}
With the generalized Euler operator the Casimir-operators are:
\begin{equation}
\Lambda_k^2 = + \hat{X}^i \hat{X}_i \hat{P}^j \hat{P}_j + 
 \hbar^2 \left(  c(\alpha)\,(k-2) J_e^k + J_e^k J_e^k \right)
\end{equation}
Now we define a Hilbert space $\mathfrak{H}_\alpha$ of all homogeneous functions $f$, which satisfy
the Laplace equation $\bar{D}^i\bar{D}_i f = 0$ and are normalized in the interval $[-1,1]$:
\begin{equation}
\mathfrak{H}_\alpha = \{ f:f(\lambda \vec{x}) = \lambda^{n \alpha}f(x);\; \bar{D}^i \bar{D}_i f = 0\} \qquad n \in \mathbb{N}
\end{equation}
This is the quantization condition. It guarantees, that solutions are regular at the origin. 

On this Hilbert space, the generalized  Euler operator $J_e^k(\alpha) $ is diagonal
and has the eigenvalues
\begin{equation}
l_k(\alpha,n)  =
\begin{cases}
0                                                  &  \text{for} \,\, n=0  \\
\frac{\displaystyle 1}{\displaystyle \Gamma(\alpha+1)}\frac{\displaystyle  \Gamma(n\alpha+1)}{\displaystyle  \Gamma((n-1)\alpha+1)}   &  \text{for} \,\, n=1,2,3,...\\
\end{cases}
\end{equation}
This is the main result of our derivation. 

We want to emphasize, that these eigenvalues are different from the degree
of homogenity in the general case $\alpha \neq 1$, or, in other words: only in the case of $\alpha=1$ the
homogenity degree $n$ of the polynoms considered coincides with the eigenvalues of 
$J_e^k(\alpha=1,n)$. 

Once the eigenvalues of the generalized Euler operator are known, 
the eigenvalues of the Casimir-operators $\Lambda_2, \Lambda_k^2$
are known, too:
\begin{eqnarray}
\Lambda_2 f & = & \hbar l_2(\alpha,n)   f \\
\Lambda_k^2 f & = & \hbar^2 l_k(\alpha,n)  (l_k(\alpha,n)   + c(\alpha)\,(k - 2)) f
\end{eqnarray}
with
\begin{equation}
l_k  \geq
l_{k-1}\geq...\geq \mid \pm l_2  \mid \geq 0
\end{equation}
For the case of only one particle ($N=1$), we can introduce the quantum numbers
j and m, which now denote the j-th or m-th eigenvalue of the Euler operator.
The eigenfunctions are fully determined by these two quantum numbers $f = \mid \! jm\!>$ 

With the definitions $L_z = L_{12}$ and $J^2 = L_{12}^2 + L_{13}^2 + L_{23}^2 $ 
it follows
\begin{eqnarray}
\label{eqLz}
L_z \mid\! jm\!> & = & \hbar l_2(\alpha,m)   \mid \!jm\!>  \\
             & &    \qquad\qquad m=0,\pm 1,\pm 2,...,\pm j \nonumber \\
\label{eqJ2}
J^2 \mid\! jm\!> & = & \hbar^2 l_3(\alpha,j)  \left( l_3(\alpha,j)+ c(\alpha) \right)  \mid \!jm\!> \\
             & & \qquad\qquad  j=0,+1,+2,... \nonumber
\end{eqnarray}
Please note the fact that $L_z$ remains unchanged for any choice of constant $c(\alpha)$, only $J^2$ changes. 

In table {\ref{tabone}} the first seven eigenvalues of $L_z$ and $J^2$ for a single particle are listed 
for $\alpha=1$, $\alpha=2/3, 0.68$ and $\alpha=0.65$ and different approximations for $c(\alpha)$. 
For $\alpha \neq 1$ the eigenvalues of the 
generalized Euler operator are not 
equally spaced any more. For $\alpha<1 $ the stepsize is strongly reduced. Since the generalized Euler operator
eigenvalues contribute quadratically into the definition $J^2$, the energy of 
higher total angular momenta is reduced increasingly.

We have derived the full spectrum of the angular momentum operator for the
fractional derivative operator Schr\"odinger type wave equation by use of 
standard algebraic methods. 

We will get additional information about the properties of this wave equation, if
we consider its factorized pendant. We present some results in the next 
section.  

\section{Results for the factorization of a non relativistic second order differential equation}
Linearization of a relativistic second order wave equation 
was first considered by Dirac{\cite{dirac}}. Starting with the relativistic
Klein-Gordon equation his derived Dirac equation gave a correct description
of the spin and the magnetic moment of the electron.

The concept of linearization is important, since it provides a well defined 
mechanism to add an additional SU(2) symmetry to a given set of symmetry
properties of a second order wave equation.

Since linearization may be interpreted as a special case of factorization, namely to
2 factors, a natural generalization is a factorization to n factors.

In 2000, Raspini\cite{raspini} proposed a Dirac-like equation with fractional derivatives of order 2/3 and found the corresponding matrix algebra 
to be related to generalized Clifford algebras;
 in 2002 Z$\acute{\text{a}}$vada \cite{zavada} generalized Dirac's approach, and found, that
relativistic covariant equations generated by taking the n-th root of the
Klein-Gordon or 
d'Alembert operator ( $\square^{1/n}$) are fractional wave equations with an
additional SU(n) symmetry.

These results indicate, that fractional order wave equations may be appropriate candidates for a description
of particles, which own a SU(n) symmetry. The case $n=3$, which corresponds to a triple factorization is
therefore important for a description of particles with a SU(3) symmetry.
 
Whether or not a factorization of non relativistic wave equations leads to similar results, has not been 
examined yet. 

In 1967, Levy-Leblond{\cite{ll}} has linearized the non relativistic 
Schr\"odinger equation and obtained a linear wave equation with an additional 
SU(2) symmetry, but until now his approach has not been extended to higher fractional order.

In order to obtain additional information on the inherent symmetries of the
fractional Schr\"odinger equation, which we proposed in (\ref{schroedinger}),
we therefore derive in the
following section  the explicit form of a fractional operator, which evolves from
a triple factorization of the ordinary Schr\"odinger equation: 

\subsection{Triple factorization of the non relativistic Schr\"odinger equation}
 
We intend to derive a fractional operator $R$, which, iterated 3 times, conforms with the ordinary, non relativistic
Schr\"odinger operator:
\begin{equation}
R R' R'' = \left( -\frac{\hbar^2}{2 m} \Delta - i \hbar \partial_t \right) \boldsymbol{1}_n 
\end{equation}
where  $\boldsymbol{1}_n$ is the $n \times n$ unit matrix.

We use the following ansatz:
\begin{eqnarray}
\label{rrr}
R &=& a A \partial^{\alpha_t}_t + b B^i \partial^{\alpha_i}_i + c C\\
R' &=& a A' \partial^{\alpha_t}_t + b B'^i \partial^{\alpha_i}_i + c C'\\
R'' &=& a A'' \partial^{\alpha_t}_t + b B''^i \partial^{\alpha_i}_i + c C''
\end{eqnarray}
with matrices $A,A',A'', B,B',B'', C,C',C''$, fractional derivative coefficients for time and
space derivative $\alpha_t, \alpha_i$ and scalar factors $a,b,c$, which will be
determined in the following.
According to Z$\acute{\text{a}}$vada \cite{zavada}, we define a triad of 
unitary, traceless $3 \times 3$ Pauli type matrices, which span a subspace of $SU(3)$
with
\begin{equation}
x_k = \exp({\frac{2 \pi i}{3}\, k }) \qquad  k = 1,2,3
\end{equation}
an explicit representation is
\begin{eqnarray}
\sigma^1 &=& 
\begin{pmatrix}
0   & x_1 & 0   \\
0   &   0 & x_2 \\
x_3 &   0 & 0   \\
\end{pmatrix}
\\
\sigma^2 &=& 
\begin{pmatrix}
0   & x_2 & 0   \\
0   &   0 & x_1 \\
x_3 &   0 & 0   \\
\end{pmatrix}
\\
\sigma^3 &=& 
\begin{pmatrix}
x_1 &   0 & 0   \\
0   &  x_2& 0   \\
0   &   0 & x_3 \\
\end{pmatrix}
\end{eqnarray}
These matrices obey an extended Clifford algebra 
\begin{equation}
\sum_{\text{all Permutations}}  \sigma^i \sigma^j \sigma^k = 6 \, \delta^{i j k} \quad i,j,k = 1,2,3 
\end{equation} 
Let $\otimes$ denote the outer product of any two
matrices. In order to describe a single particle with the coordinates $\{ t,x,y,z\} $, we define the following 4 $\gamma$ matrices,
with dimension $9 \times 9$:
\begin{eqnarray}
\gamma^0 &=& \boldsymbol{1}_3 \otimes \sigma^3 \\
\gamma^i &=& \sigma^i \otimes \sigma^1  \qquad i = 1,2,3
\end{eqnarray}
Now we are able to specify the above introduced matrices:
\begin{eqnarray}
A  &=& \frac{1}{\sqrt{3}} (\gamma^0 - x_1 \, \boldsymbol{1}_9) \\
A' &=& \frac{1}{\sqrt{3}} (\gamma^0 - x_2 \, \boldsymbol{1}_9) \\
A''&=& \frac{1}{\sqrt{3}} (\gamma^0 - x_3 \, \boldsymbol{1}_9) \\
B^i &=& \gamma^i  \\
B'^i &=& \gamma^i  \\
B''^i &=& \gamma^i  \\
C &=& x_1 \, \boldsymbol{1}_3 \otimes 
\begin{pmatrix}
1   & 0 & 0   \\
0   & 0 & 0   \\
0   & 0 & 0   \\
\end{pmatrix}
\\
C' &=& x_2 \, \boldsymbol{1}_3 \otimes 
\begin{pmatrix}
0   & 0 & 0   \\
0   & 1 & 0   \\
0   & 0 & 0   \\
\end{pmatrix}
\\
C'' &=& x_3 \, \boldsymbol{1}_3 \otimes 
\begin{pmatrix}
0   & 0 & 0   \\
0   & 0 & 0   \\
0   & 0 & 1   \\
\end{pmatrix}
\end{eqnarray}
with these specifications $R \, R' R''$ yields
\begin{equation}
R \, R' R'' = \left( a^2 c \, \partial^{2 \alpha_t}_t + b^3 \sum_{i=1}^3 \partial^{3 \alpha_i}_i \right) \boldsymbol{1}_9
\end{equation}
A term by term comparison with the nonrelativistic Schr\"odinger operator determines the 
fractional derivative coefficients: 
\begin{eqnarray}
\alpha_t &=& 1/2\\
\alpha_i &=& 2/3
\end{eqnarray}
and the scalar factors:
\begin{eqnarray}
a &=& (-i \hbar)^{1/2} \left(  \frac{1}{m c^2}  \right)  ^{1/6}\\
b &=& - \left( \frac{\hbar^2}{2 m} \right) ^{1/3}\\
c &=& (m c^2)^{1/3}
\end{eqnarray}
Finally, according to (\ref{defs}) and (\ref{caputo}), we extend the derivative operator on $\mathbb{R}$ via:
\begin{equation}
\partial_\mu \rightarrow \text{sign}(x^\mu) \, ^{+}_{c}D(|x^\mu|)
\end{equation}
Thus the fractional operators $R, \, R', R'' $ are completely determined. As a remarkable fact we note the 
different fractional derivative coefficients for the fractional time and space derivative. 

We therefore have proven, that the
resulting SU(3) symmetry is neither a consequence of a relativistic treatment nor is it a specific
property of linearization (which means, specific to first order derivatives of time and space respectively). 

It is
a consequence of the triple factorization solely.

We want to emphasize, that the factorization not only determines the symmetry group
of the $\gamma$-matrices used,  but also determines the dynamics of the system, forcing e.g. for a SU(3)
symmetry  fractional
time ($\alpha_t = 1/2$) and space derivatives($\alpha_i = 2/3$). 

Thus, the result of factorization shows two important properties: 
It yields an additional SU(n) symmetry and simultaneously the
 corresponding dynamics.
Applied to QCD, this seems more consistent than the standard  concept based on Yang-Mills field theories, where 
an arbitrary non abelian
 symmetry group gauge field, which first has to be deduced from experimental data to be a SU(3) symmetry
is coupled to a symmetry 
independent
dynamical (e. g. Dirac) field, neglecting a possible influence of the symmetry onto the dynamics.
   
In that sense, the fractional operator $R$, defined in (\ref{rrr}), would be an alternative starting point for a pure, non relativistic QCD, since
it contains a consistent description of both, symmetry and dynamics of a pure SU(3) symmetry, without any 
additional SU(2) admixture.

Finally, besides the fractional wave equation operator $R$ and the triple iterated $R \, R' R''$, which corresponds
to the ordinary Schr\"odinger operator, an additional type of wave equation, the twofold iterated
$R\, R'$ emerges, which reads:
\begin{equation}
c R\, R' =  a^2 c A A'  \partial_t   + b^2 c B^i B^j \partial^{2/3}_i \partial^{2/3}_j + \text{additional terms}
\end{equation}
or, inserting the factors:
\begin{eqnarray}
\label{s2}
S^{(2)} &=&  -i \hbar  A A'  \partial_t   - \frac{1}{2}(\frac{\hbar}{m c})^{4/3} m c^2 \,
  (\frac{1}{2})^{1/3}\,B^i B^j \partial^{2/3}_i \partial^{2/3}_j \nonumber \\
& & + \text{additional terms}
\end{eqnarray}
Obviously  $S^{(2)}$ and  the fractional Schr\"odinger equation ({\ref{schroedinger}}), we derived in section {\ref{sgl}}, are 
closely related.

$H^\alpha$ seems nothing else but a 
scalar version of $S^{(2)}$ for the special case $\alpha=2/3$. Therefore, examination of 
the properties of the scalar 
fractional Schr\"odinger equation ({\ref{schroedinger}}) with $\alpha=2/3$ should reveal some properties of an 
inherent $SU(3)$ symmetry. 

Summarizing all facts collected, we assume, that the fractional derivative operator
Schr\"odinger type wave equation with $\alpha=2/3$ is an appropriate candidate for a 
non relativistic description of particles with quark-like properties.


\section{Interpretation of the Charmonium spectrum}
In the previous sections we have introduced the concept of fractional derivative
operators and discussed some properties of the resulting non relativistic fractional Schr\"odinger type  
equation and its factorized pendant.

\begin{figure}
\begin{center}
\includegraphics[width=75mm,height=85mm]{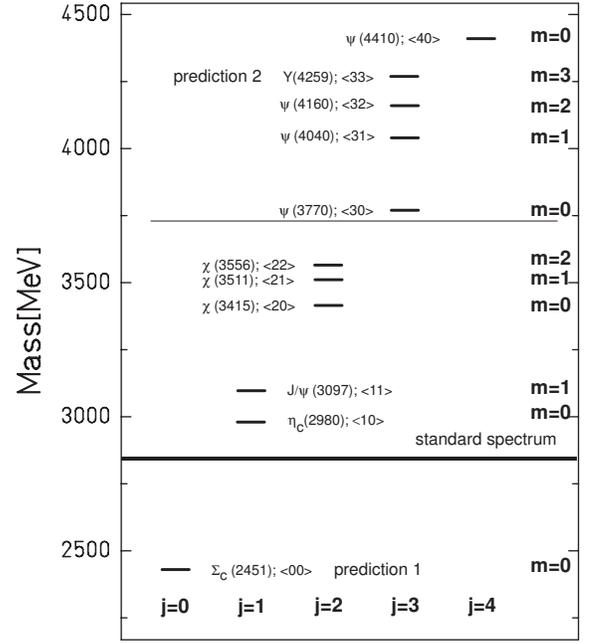}\\
\caption{\label{spectrum}Charmonium spectrum. All observed particles are given with their name, experimental
mass from{\cite{pdg}} and the proposed $SO^\alpha(3)$ conforming quantum numbers j and m, which  are the j-th and m-th eigenvalue of 
the generalized Euler operator. Predicted particles are: in the lower part of the spectrum denoted as prediction 1 
$<\!00\!>$ associated with $\Sigma_c^0(2455)$ and in the upper part prediction 2 $<\!33\!>$ associated with Y(4260).    
} 
\end{center}
\end{figure}

We have developed a new theoretical concept, which fulfills at least the following
three demands:
First, available experimental data will be reproduced with a reasonable accuracy.
Second, it will give new insights on underlying symmetries and properties
of the objects under consideration. 
Third, we will make predictions, which can be proven by experiment.

None of our results, presented so far, does require any information from QCD or a similar
theory. All our statements could have been made in the 1930s already, even though they
would have been highly speculative. Today, we are in the comfortable position, that
there are enough experimental data, our predictions can be compared with.

A promising candidate is the charmonium spectrum{\cite{Ei75}}. 

In the upper part of figure {\ref{spectrum}}
we have displayed all experimentally observed charmonium-states with experimental masses, 
which are normally compared with results from a potential model, which tries to simulate
confinement and attraction by fitting a model potential{\cite{Ei75}},{\cite{Kr79}}.

We will assume, that this spectrum is a single particle spectrum for a 
particle, whose properties are described by the free fractional  
Schr\"odinger type equation({\ref{schroedinger}}). We suppose, the system is rotating in a minimally
coupled field, which causes a magnetic field $B_j$. This leads to the following
 Hamiltonian H or mass formula
\begin{equation}
\label{massformula1}
H(j,m) \mid\! jm\!> = \left( \kappa J^2/\hbar^2 + B_j L_z/\hbar + m_0 c^2 \right) \mid \!jm\!> 
\end{equation}
where  $\kappa$, $B_j$ and $m_0 c^2$ will be adjusted to the experimental data.

The eigenfunctions $\mid \!jm\!>$ 
are modified spherical harmonics with $D^i D_i \mid\! jm\!> =0$, the eigenvalues for $J^2$ and $L_z$
are given by ({\ref{eqLz}}),({\ref{eqJ2}}) and are listed in table 1. 

A first remarkable observation is the fact, that states with $L_z < 0$ are missing in the experimental 
spectrum. Only right-handed particles are realized. 

This may be due to the fact, that actually $L_z^2$ is
a Casimir-Operator of $SO^\alpha(2)$, while $L_z$ is not and therefore the multipletts should more precisely 
be classified
according to the
$m^2$ quantum number in the general case $\alpha \neq 1$.  Consequently in the following we will work with
$L_z \geq 0$. 

To check the influence of our approximations ({\ref{cp1}}),({\ref{cp2}}) and ({\ref{cp3}}) of $c(\alpha)$ given in ({\ref{comm}}) we will proceed in two steps: 
First, we will consider the case $c_0 = 1$. 
In a second step, we will test the influence of the successive 
approximations for $c(\alpha)$ on the accuracy of the proposed mass formula with least square fits on the 
charmonium spectrum.

\subsection{Interpretation of the charmonium spectrum in the case $c_0=1$}

We first consider the case $c(\alpha) = 1$. 
The corresponding $J^2$ and $L_z$ values used are listed in table {\ref{tabone}} as $L_z(\alpha)$
and $J^2_0(\alpha)$. 

The first crucial test is the verification of the correct value of the non trivial $m=2$ quantum number
which corresponds to the $n=2$ eigenvalue of the generalized Euler operator. 

For the set of $\chi$-particles ($j=2$) (experimental masses and errors are taken from {\cite{pdg}}), we obtain:
\begin{eqnarray}
L_z(j=2,m=2)_{\text{exp}}&=&\frac{\chi(3556)\!<\!22\!> -\chi(3415)\!<\!20\!>}{\chi(3511)\!<\!21\!> -\chi(3415)\!<\!20\!>} \nonumber \\
& =&   1.478 \pm 0.007
\end{eqnarray}
Thus, an $\alpha$ from the experimental spectrum is deduced:
\begin{equation}
\alpha_{\text{exp}}^{j=2} = 0.680 \pm 0.006
\end{equation}
Which is remarkably close to the theoretically expected $\alpha_{\text{th}} = 2/3$.

An alternative approach to determine the experimental value for $\alpha$ or the $m=2$ quantum number repectively,
follows from  the set of $\Psi$-particles ($j=3$):
\begin{eqnarray}
L_z(j=3,m=2)_{\text{exp}}&=& \frac{\Psi(4160)\!<\!32\!> -\Psi(3770)\!<\!30\!>}{\Psi(4040)\!<\!31\!> -\Psi(3770)\!<\!30\!>} \nonumber \\
& =&  1.44 \pm 0.09 
\end{eqnarray}
A second  experimental $\alpha$ is obtained:
\begin{equation}
\alpha_{\text{exp}}^{j=3} = 0.65 \pm 0.08
\end{equation}

Within experimental errors, both values are identical. This observation supports the assumption, that the 
spectrum may be interpreted using one unique $\alpha$.

According to our level scheme, the $\Psi\!<\!33\!>\!$ state is missing in the standard charmonium spectrum. Using
$\alpha_{\text{exp}}^{j=2} = 0.680$ and $L_z(\alpha,m)$ from table {\ref{tabone}} the predicted mass is:
\begin{eqnarray}
\Psi\!\!<\!\!33\!\!>\!&=& \frac{L_z(\alpha,m=3)}{L_z(\alpha,m=2)}( \Psi(4160)\!\!<\!\!32\!\!> -\Psi(3770)\!\!<\!30\!\!>) \nonumber \\
& =&  4268 \pm 22 [\text{MeV}] 
\end{eqnarray}
In june 2005, the Babar collaboration announced the discovery of a new charmonium state named Y(4260) {\cite{aubert}}. 
The reported mass of 4259$[\text{MeV}]$ is in excellent agreement with our mass prediction for $\Psi\!<\!33\!>$.  Therefore,
we associate the predicted particle with $Y(4260)$.      

Next we will determine the constants $m_0 c^2$ and $\kappa$ in ({\ref{massformula1}}). 

We choose the two lowest experimental states of the
standard charmonium spectrum, $\eta_c(2980)\!<\!10\!>$ and $\chi(3415)\!<\!20\!>$. With $\alpha_{\text{exp}}^{j=2} = 0.680$ 
and $J^2_0(\alpha,j)$ from table {\ref{tabone}}
we obtain a set of equations
\begin{eqnarray}
m_0 c^2 + 2 \kappa         &=& \eta_c(2980)\!<\!10\!>\\
m_0 c^2 + 3.663108 \kappa  &=& \chi(3415)\!<\!20\!> 
\end{eqnarray}
which determine 
\begin{eqnarray}
\label{mc2}
m_0 c^2 &=& 2455 \pm 3[\text{MeV}] \nonumber \\
\kappa  &=& 262.4 \pm 0.9[\text{MeV}] 
\end{eqnarray}

Our level scheme predicts a particle with quantum numbers $<\!00\!>$, which is beyond the scope of 
charmonium potential models. According to our mass formula, it has a 
predicted mass of $<\!00\!>= 2455 \pm 3[\text{MeV}]$. 

Since this is a low lying state, it should already have been observed.
Indeed, there exists an appropriate candidate, the 
$\Sigma_c^0(2455)\!<\!00\!>$ baryon, with an experimental mass of $2452.2 [\text{MeV}]$. 
This is a charmed baryon with quark content (ddc). 

The minimal difference of only $2.8[\text{MeV}]$
between predicted and experimental mass of the $\Sigma_c^0(2455)\!<\!00\!>$ particle indicates, that the
assumed fractional $SO^\alpha(3)$ symmetry is fulfilled exactly. 

Obviously, the 
fractional $SO^\alpha(3)$ multipletts describe mesonic and baryonic states of the charm-quark simultaneously.

Due to its experimentally observed properties, the internal structure of the $Y(4260)\!\!<\!\!33\!\!>$ particle is subject of 
actual discussion{\cite{zhu}}. Besides being a conventional $c \bar{c}$ state, it could alternatively
be a tetraquark with constituents $(u\bar{u}c\bar{c})$ or a hybrid charmonium. 

Finally we have only one experimental candidate for $j=3$ and $j=4$ respectively.
With parameters (\ref{mc2}) using the mass formula (\ref{massformula1}) we obtain the theoretical values 
\begin{eqnarray}
\label{j3}
\Psi(3770)\!<\!30\!>_{th} &=& 3894 \pm 8 [\text{MeV}]\\
\label{j4}
\Psi(4410)\!<\!40\!>_{th} &=& 4406 \pm 10 [\text{MeV}]
\end{eqnarray}
For $j=3$ the calculated mass (\ref{j3}) differs by $124[MeV]$ from the experimental value. 

On the other hand, the theoretical $\Psi(4410)\!<\!40\!>$ mass  (\ref{j4}) 
matches exactly with the experimental value within the experimental errors.

This indicates, that the particles for $j=3$, observed in experiment, carry an additional property, which 
reduces the mass by the amount of e.g. a pion. Of course,
if we add an additional $(\Delta\tau) \delta_{j3}$ term to the proposed mass formula, we can shift these levels 
by the necessary amount. 

Summarizing these results, the charmonium spectrum reveals an underlying $SO^\alpha(3)$ symmetry, which agrees
with the predictions of our theory in the case of $\alpha \approx 2/3$. The eigenvalues of the generalized
Euler operator conform within experimental errors with experimental data.
Extending the standard charmonium spectrum, two additional particles
have been predicted and associated with $\Sigma_c^0(2455)$ and $Y(4260)$ observed recently.

\subsection{Least square fits  of the charmonium spectrum}
\begin{table*}
\caption{\label{tabmass}Optimum parameter sets for a fit of the experimental charmonium spectrum with 
mass formula ({\ref{massformula3}}) in units $[$MeV$]$. 
Errors $\Delta m$ are given for the subset $m=0$ and the full spectrum $\Delta m_{all}$. The first row
corresponds to $\alpha=2/3$ fixed, the following three rows  correspond 
to the three approximations for $c_0$,$c_1$ and $c_2$ according to 
({\ref{cp1}}),({\ref{cp2}}) and ({\ref{cp3}}) with $\alpha$ optimized. In table {\ref{tabcomp}} resulting
theoretical masses and errors are listed.}
\begin{ruledtabular}
\begin{tabular}{@{}*{11}{r}}
$\alpha$   & $c(\alpha)$  &$m_0c^2$&$\kappa$& $B_1$  &$B_2$ &$B_3$ &$\Delta \tau$& $\Delta m_{m=0}$ & $\Delta m_{all}$&comment\cr
2/3     & 1.00  & 2439.33 & 274.66 & 108.25 & 87.00 & 263.69& -129.04 &  5.68&  8.98 &$\alpha$ fixed \cr
0.681   & 1.00  & 2451.26 & 263.83 & 117.98 & 93.39 & 259.16& -124.48 &  1.86&  2.02          & $\alpha$ variation\cr
0.647   & 0.545 & 2452.67 & 336.16 & 119.79 & 95.72 & 270.19&-129.00 &  1.14 &  1.15 &$\alpha$ variation\cr
0.649   & c(j,$\alpha$)& 2451.90 & 367.41 & 116.13 &  98.37& 269.46&-124.39 &  0.73&  0.79 &$\alpha$ variation\cr
\end{tabular}
\end{ruledtabular}
\end{table*}

In the previous section we gave an interpretation of the charmonium spectrum for the case $c(\alpha)=1$. We found,
that the spectrum may be described quantitatively, using the proposed  mass formula  for $\alpha=0.680$. 

Extending the mass formula ({\ref{massformula1}}) including a correction term for the $j=3$ multiplett, we use
\begin{equation}
\label{massformula3}
H(j,m) \! \mid \! jm\!> =  \kappa J^2/\hbar^2 + B_j L_z/\hbar+m_0 c^2 + \delta_{3\,j} \Delta \tau 
\! \mid \! jm\! > 
\end{equation}
to find a fit on the experimental charmonium spectrum. 

To prove, that for $c_0=1$,  $\alpha=0.68$ indeed is the appropriate
choice for an interpretation of the charmonium spectrum, we minimized errors with respect to $\alpha$ and obtained 
$\alpha=0.681$. In table {\ref{tabmass}} the optimum parameter sets for $\alpha=2/3$ and $c_0, c_1, c_2$
 and resulting errors 
are tabulated. 

For $c_1(\alpha)$ from ({\ref{cp2}}), $\alpha=0.68$ is not the optimum choice any more.
 We therefore minimized errors with respect to $\alpha$, finding
$\alpha=0.647$ for this case. 

For $c_2(j,\alpha)$ from ({\ref{cp3}}), we observe a minimal shift in $\alpha$, finding
$\alpha=0.649$ for this case. This indicates, that a more sophisticated treatment of $c(\alpha)$ 
will only cause neglible changes for $\alpha$ and corresponding parameter sets.

Comparing the optimum parameter sets, the changes in the treatment of $c(\alpha)$ are mainly
absorbed by the parameter $\kappa$. Parameter $\Delta \tau$ remains remarkably constant. This supports
interpretation for this parameter to be a $j=3$ specific quantum number. 

A comparison of experimental with calculated masses, based on the optimum parameter sets, is given in table {\ref{tabcomp}}. 
Mass differences are less than $0.1 \%$ and decreasing for $c_0,c_1,c_2$. 

With the optimum parameter sets we can predict the mass of the $<50>$ state to be
\begin{equation}
X\!<\!50\!>_{\text{th}} = 4965 \pm 10 [\text{MeV}]
\end{equation}
This state has not been observed in experiments yet.

We conclude, that the proposed $SO^\alpha(3)$ symmetry is fulfilled
exactly. The values of $\alpha = 0.68$ and $\alpha = 0.65$ resulting from the least square fits
are close to , but differ significantly from the theoretically expected $\alpha=2/3$. 
This indicates, that the inherent $SU(3)$ symmetry is almost fulfilled
exactly, with a difference of only $2\%$. 
\begin{table*}
\caption{\label{tabcomp}Comparison of experimental and calculated masses according to mass formula
({\ref{massformula3}}) with optimized parameter sets listed in table  {\ref{tabmass}} in units $[\text{MeV}]$.
The last row lists predicted theoretical masses for $X\!<\!\!50\!\!>$. }

\begin{ruledtabular}
\begin{tabular}{@{}*{11}{r}}
       &                 &           &\multicolumn{2}{c}{$\alpha=2/3,\, c_0=1$}   &\multicolumn{2}{c}{$\alpha=0.68,\, c_0=1$}   &\multicolumn{2}{c}{$\alpha=0.647,\, c_1(\alpha)$}&\multicolumn{2}{c}{$\alpha=0.649,\, c_2(j,\alpha)$}\cr
$<\!jm\!>$ & symbol          & $m_{exp}$ & $m_{th}$& $\Delta m_{th}$ & $m_{th}$& $\Delta m_{th}$ & $m_{th}$&$\Delta m_{th}$ & $m_{th}$&$\Delta m_{th}$ \cr
$<\!00\!>$ & $\Sigma_c^0$    & 2452.2    &  2439.33& -12.87 & 2451.26 &  -0.93          & 2452.67 &   0.47 & 2451.90  &   -0.30 \cr
$<\!10\!>$ & $\eta_c$        & 2979.6    &  2988.66&  9.05  & 2978.94 &  -0.66          & 2977.12 &  -2.47 & 2980.78  &    1.18 \cr
$<\!11\!>$ & $J/\Psi$        & 3096.9    &  3096.92&  0 &               3096.92 &   0             & 3096.92 &   0    & 3096.92  &    0    \cr
$<\!20\!>$ & $\chi_0$        & 3415.2    &  3426.89&  11.70&            3417.80 &   2.60          & 3417.15 &   1.95 & 3413.65  &    -1.54\cr
$<\!21\!>$ & $\chi_1$        & 3510.6    &  3513.89&   3.30&           3511.19 &   0.60          & 3512.87 &   2.27 & 3512.03  &     1.43\cr
$<\!22\!>$ & $\chi_2$        & 3556.3    &  3554.00&  -2.26&            3555.85 &  -0.40          & 3554.68 &  -1.58 & 3555.26  &     1.00\cr
$<\!30\!>$ & $\Psi$          & 3770      &  3772.35&  2.34&             3773.92 &   3.92          & 3770.17 &   0.17 & 3770.41  &     0.41\cr
$<\!31\!>$ & $\Psi$          & 4040      &  4036.04&  -3.95&            4033.08 &  -6.91          & 4040.37 &   0.37 & 4039.87  &    -0.13\cr
$<\!32\!>$ & $\Psi$          & 4160      &  4157.60&  -2.39&            4157.02 &  -2.98          & 4158.39 &  -1.61 & 4158.30  &    -1.70\cr
$<\!33\!>$ & $Y(4260)$       & 4259      &  4263.01&  4.01&            4264.98 &   5.97          & 4260.08 &   1.07 & 4260.42  &     1.42\cr
$<\!40\!>$ & $\Psi$          & 4415      &  4406.07&   -8.92&           4413.80 &  -1.20          & 4414.36 &  -0.64 & 4415.22  &     0.22\cr
$<\!50\!>$ & $X$             &           &  4937.06&         &          4959.54 &                 & 4957.54 &        & 4969.07  &     \cr
\end{tabular}
\end{ruledtabular}
\end{table*}

\subsection{Size estimate for $\Sigma_c^0$}
Up to now we have treated $m_0 c^2$ as a simple parameter of the proposed mass formula. As a result of
our discussion above, we associate $H(0,0) = m_0 c^2$ with the mass of $\Sigma_c^0$. If $H(j,m)$ was the
solution of the Schr\"odinger equation with a given potential V, then $H(0,0)$ would be interpreted as the
zero point energy in this potential plus the rest mass of its constituents.

We intend to estimate the size of $\Sigma_c^0$. Therefore we will calculate the expectation value $<\! \hat{r}\!>$
of
the radius operator $\hat{r}$, which is given by
\begin{eqnarray}
\hat{r} &=& \sqrt{\hat{X}_1^2+\hat{X}_2^2+ \hat{X}_3^2}\\
        &=& \left( \frac{\hbar}{m c} \right)^{1-\alpha} \! \!  \frac{1}{\Gamma(1 + \alpha)} \sqrt{x_1^{2 \alpha}+x_2^{2 \alpha}+x_3^{2 \alpha}} 
\end{eqnarray}
For a first estimate, we choose the infinite square well potential(\ref{potsquare}). The energy eigenvalues
are given by (\ref{energy_squarewell}).
Therefore
\begin{eqnarray}
\Sigma_c^0 &=& (2 m_d  + m_c) c^2 + E_0(N=3,\alpha)\\
&= &  (2 m_d  + m_c)  c^2 + \frac{3}{2} \left( \frac{\hbar}{m_c c}\right)^{2 \alpha} m_c c^2 |\frac{k^0_0}{a}|^{2 \alpha}\nonumber
\end{eqnarray}
determines the half boxsize $a$. 
The wave function was defined in (\ref{defcossin}).  
Therefore with the abbreviations
\begin{eqnarray}
d V^\alpha &=& dx_1^\alpha dx_2^\alpha dx_3^\alpha \\
\Psi(x_1,x_2,x_3) &=& \cos(k^0_0 x_1/a)\cos(k^0_0 x_2/a) \cos(k^0_0 x_3/a) \nonumber  
\end{eqnarray}
the expectation value according to ({\ref{expect}}) is
\begin{equation}
\label{expectR}
<\!\hat{r}(\Sigma_c^0)\!> = \frac{\iiint_0^a dV^\alpha \Psi^* \hat{r} \Psi}{\iiint_0^a dV^\alpha \Psi^* \Psi}
\end{equation}
Setting $\alpha=2/3$ and
\begin{eqnarray}
\Sigma_c^0  &=& 2452.2 [\text{MeV}] \\
m_d c^2     &=& 300  [\text{MeV}]  \nonumber \\
m_c c^2     &=& 1400  [\text{MeV}]  \nonumber \\
k^0_0       &=& 1.1648 \, \pi/2  \nonumber 
\end{eqnarray}
we derive $a = 0.81[\text{fm}]$ and therefore we obtain the expectation value for the radius
\begin{equation}
<\!\hat{r}(\Sigma_c^0)\!>_{\text{cube}} = 0.32 [\text{fm}] 
\end{equation}
Similarly, we can proceed for the  infinite spherical well potential (\ref{potsphere}).

For the  spherical ground state wave function
in carthesian coordinates(\ref{radx}), we obtain
\begin{eqnarray}
L_z \, g(N=3,\alpha,x_1,x_2,x_3)  &=& 0 \\
J^2 \, g(N=3,\alpha,x_1,x_2,x_3)  &=& 0  
\end{eqnarray}
and therefore $g$ indeed is the ground state $|00\!\!>$.
 
The ground state energy $e_0$ is given by (\ref{e0})
\begin{eqnarray}
\Sigma_c^0 &=& (2 m_d  + m_c) c^2 + e_0(N=3,\alpha)\\
&= &  (2 m_d  + m_c) c^2 + \frac{1}{2} \left( \frac{\hbar}{m_c c}\right)^{2 \alpha} \! \! m_c c^2 |\frac{k^0_\text{sph}}{r_0}|^{2 \alpha}\nonumber
\end{eqnarray}
with $k^0_\text{sph}=3.1652 \, \pi/2  $ this determines $r_0 = 1.08[\text{fm}]$. 

With (\ref{expectR}) we obtain for the expectation value of the radius
\begin{equation}
<\!\hat{r}(\Sigma_c^0)\!>_{\text{sphere}} = 0.33 [\text{fm}] 
\end{equation}
Consequently, both potentials lead to similar expectation values.

Since $\Sigma_c^0$ is not within the scope of standard charmonium models, there is no direct comparison.

Nevertheless,
there are  radii, derived from charmonium model calculations{\cite{Ei75}}, reported for {\cite{gerland1}}-{\cite{gerland3}}:
\begin{eqnarray}
<\!\!\hat{r}({J/\psi\!<\!11\!>\!})\!\!> &\approx& 0.2 [\text{fm}]\\ 
<\!\!\hat{r}({\chi_0\!<\!20\!>\!})\!\!> &\approx& 0.3 [\text{fm}]\\
<\!\!\hat{r}(\Psi   \!<\!30\!>\!)\!\!>   &\approx& 0.4 [\text{fm}]
\end{eqnarray}

Therefore our results are reasonable compared with these calculations.

\section{Conclusion}
Based on the Caputo fractional derivative, we have defined a fractional derivative operator for arbitrary 
fractional order $\alpha$. 
A Schr\"odinger type wave equation,
derived by quantization of the classical non relativistic Hamiltonian, generates  free particle solutions, 
which are
confined to a certain region of space. Therefore confinement is a natural consequence of the use of a fractional 
wave equation. 

The multiplets of the generalized angular momentum operator have been  classified
acoording to the $SO^\alpha(3)$ scheme, the spectrum of the Casimir-Operators has been calculated analytically. 

We have also
shown, that for $\alpha=2/3$, corresponding to a fractional non relativistic Levy-Leblond wave function an inherent SU(3) 
symmetry is apparent. 

From a detailed discussion of the charmonium spectrum we conclude, that the spectrum may be
understood quantitatively within the framework of our theory. Approximately $\alpha \approx 2/3$ is valid. The
experimental masses are reproduced wih an accuracy better than $0.1 \%$.

Extending the standard charmonium spectrum, three new particles
have been  predicted, two of them  associated with $\Sigma_c^0(2455)$, a charmed baryon and $Y(4260)$, observed recently.
The third  particle, labeled $X\!\!<\!\!50\!\!>$  according to the proposed $SO^\alpha (3)$ level scheme, 
with a predicted mass of
$4965 \pm 10  [\text{MeV}]$, has not been experimentally verified yet.  

Summarizing the results of our considerations, the proposed fractional non relativistic Schr\"odinger type wave equation 
is a powerful alternative for a discussion of charmonium properties and extends our knowledge beyond the standard 
achieved with phenomenological
models. 

Therefore fractional wave equations may play an important role
in our understanding of particles with quark-like properties, e.g. confinement.

\section{Acknowledgements}
We thank A. Friedrich and G. Plunien from TU Dresden, Germany, for fruitful discussions.
\end{document}